\documentstyle[preprint,eqsecnum,aps]{revtex}
\begin{document}
\draft
\title{RELEVANCE OF NUCLEON SPIN IN AMPLITUDE ANALYSIS\\
OF REACTIONS $\pi^- p \to \pi^0 \pi^0 n$ AND $\pi^- p \to \eta\eta n$.}
\author{M. Svec\footnote{electronic address: svec@hep.physics.mcgill.ca}}
\address{Physics Department, Dawson College, Montreal, Quebec,  
Canada H3Z 1A4\\
and\\
McGill University, Montreal, Quebec, Canada H3A 2T8}
\maketitle
\begin{abstract}
The measurements of reactions $\pi^- p \to \pi^- \pi^+ n$ and  
$\pi^+ n \to \pi^+ \pi^- p$ on polarized targets at CERN found a  
strong dependence of pion production amplitudes on nucleon spin.  
Analyses of recent measurements of $\pi^- p \to \pi^0\pi^0 n$  
reaction on unpolarized targets by GAMS Collaboration at 38 GeV/c  
and BNL E852 Collaboration at 18 GeV/c use the assumption that pion  
production amplitudes do not depend on nucleon spin, in conflict  
with the CERN results on polarized targets. We show that  
measurements of $\pi^- p \to \pi^0\pi^0 n$ and $\pi^- p \to \eta\eta  
n$ on unpolarized targets can be analysed in a model independent  
way in terms of 4 partial-wave intensities and 3 independent  
interference phases in the mass region where $S$- and $D$-wave  
dominate. We also describe model-independent amplitude analysis of  
$\pi^- p \to \pi^0\pi^0 n$ reaction measured on polarized target,  
both in the absence and in the presence of $G$-wave amplitudes. We  
suggest that high statistics measurements of reactions $\pi^- p \to  
\pi^0 \pi^0 n$ and $\pi^- p \to \eta\eta n$ be made on polarized  
targets at Protvino IHEP and at BNL, and that model-independent  
amplitude analyses of this polarized data be performed to advance  
hadron spectroscopy on the level of spin dependent production  
amplitudes.
\end{abstract}
\pacs{}
\section{Introduction}
The dependence of hadronic reactions on nucleon spin was discovered  
by Owen Chamberlain and his group at Berkeley in 1957 in  
measurements of polarization in $pp$ and $np$ elastic scattering at  
320 MeV\cite{chamber57}. The prevalent belief in 1950's and 1960's  
was that in hadronic reactions spin is irrelevant and the spin  
effects observed by Chamberlain were expected to vanish at very high  
energies, such as 6 GeV/c. Instead, measurements of polarization in  
two body reactions found significant dependence on spin up to 300  
GeV/c at CERN\cite{fidecaro81} and Fermilab\cite{kline80}.  
Measurements at BNL found large spin effects at very large momentum  
transfers\cite{heppelman85,crab88}. Inclusively produced hyperons  
show large polarizations up to the equivalent of 2000  
GeV/c\cite{heller90}.  Large spin effects in inclusive reactions  
were observed at Fermilab Spin Facility with polarized proton and  
antiproton beams at 200 GeV/c\cite{bravar96,adams96}. Today, work is  
in progress to study dependence of hadronic reactions on spin and  
nucleon spin structure with polarized colliding proton beams at RHIC  
collider at BNL\cite{yoko95}.

The most remarkable feature of hadronic reactions is the conversion  
of kinetic energy of colliding hadrons into the matter of produced  
particles. This conversion process is characterized by conservation  
of total four momentum and quantum numbers such as electric charge,  
baryon number and strangeness. The conversion process depends also  
on the flavour content and spin of colliding hadrons.

The simplest production processes are single-pion production  
reactions such as $\pi N \to \pi^+ \pi^- N$ and $KN \to K\pi N$. In  
1978, Lutz and Rybicki showed\cite{lutz78} that measurements of  
these reactions on polarized target yield enough observables that  
model independent amplitude analysis is possible determining the  
spin dependent production amplitudes. The measurements of these  
reactions on polarized targets are thus of special interest because  
they permit to study the spin dependence of pion creation directly  
on the level of production amplitudes. Several such measurements  
were actually done at CERN-PS.

The high statistics measurement of $\pi^- p \to \pi^- \pi^+ n$ at  
17.2 GeV/c on unpolarized target\cite{grayer74} was later repeated  
with a transversely polarized proton target at the same  
energy\cite{groot78,becker79,becker79b,chabaud83,sakrejda84,rybicki85}.  
Model independent amplitude analyses were performed for various  
intervals of dimeson mass at small momentum transfers $-t =  
0.005-0.2$ (GeV/c)$^2$\cite{groot78,becker79,becker79b,chabaud83},  
and over a large interval of momentum transfer $-t = 0.2 - 1.0$  
(GeV/c)$^2$\cite{sakrejda84,rybicki85}.

Additional information was provided by the first measurement of  
$\pi^+ n \to \pi^+ \pi^- p$ and $K^+ n \to K^+ \pi^- p$ reactions on  
polarized deuteron target at 5.98 and 11.85  
GeV/c\cite{lesquen85,lesquen89}. The data allowed to study the  
$t$-evolution of mass dependence of moduli of  
amplitudes\cite{svec90}. Detailed amplitude  
analyses\cite{svec92,svec92b} determined the mass dependence of  
amplitudes at larger momentum transfers $-t=0.2-0.4$ (GeV/c)$^2$.

The crucial finding of all these measurements was the strong  
dependence of production amplitudes on nucleon spin. The process of  
pion production is very closely related to nucleon transversity, or  
nucleon spin component in direction perpendicular to the production  
plane. For instance, in $\pi^- p \to \pi^- \pi^+ n$ at small $t$ and  
dipion masses below 1000 MeV, all amplitudes with recoil nucleon  
transversity down are smaller than the transversity up amplitudes,  
irrespective of dimeson spin and helicity. All recoil nucleon  
transversity down amplitudes also show suppression of resonance  
production in the $\rho$ meson region.

The measurements of $\pi N \to \pi^+ \pi^- N$ reactions on  
polarized target also enabled a model-independent separation of $S$-  
and $P$-wave amplitudes. The $S$-wave amplitude with recoil nucleon  
transversity up is found to resonate at 750 MeV in both  
solutions\cite{svec92c,svec96,svec96b} irrespective of the method of  
amplitude analysis\cite{svec96b}. The resonance is narrow and the  
most recent fits\cite{svec96b} determined its width to be $108\pm  
53$ MeV.

Recently high statistics measurements of $\pi^- p \to \pi^0 \pi^0  
n$ reaction were made at 38 GeV/c by the GAMS Collaboration at IHEP  
Protvino\cite{alde95,kondashov95,kondashov96} and at 18 GeV/c by the  
E852 Collaboration at BNL\cite{brabson95}. In principle one expects  
these experiments to confirm the existence of $\sigma(750)$ state  
and to search for new states in higher partial waves. However the  
situation is not so simple. The reason is that both groups analyse  
their well acquired data using a strong simplifying assumption that  
the production amplitudes are independent of nucleon  
spin\cite{costa80,chung96,chung96b,kondashov,gunter}. The purpose of  
this assumption is to reduce the number of unknown amplitudes by  
one half and to enable to proceed with amplitude analysis using such  
spin independent ``amplitudes''.

At this point it is important to realize that one does not really  
make an assumption that production amplitudes are independent on  
nucleon spin. It is a well-known fact that nucleon helicity nonflip  
and flip amplitudes have entirely different $t$-dependence due to  
conservation of angular momentum. The helicity flip amplitudes  
vanish as $t\to 0$ while helicity nonflip amplitudes do not. The  
model independent amplitude analyses of two-body reactions also  
found that the zero structure of flip and nonflip amplitudes are  
dramatically different. Moreover, the pion production at small $t$  
proceeds mostly via the pion exchange which contributes to helicity  
flip amplitudes. Thus the assumption that is really being made is  
that all nonflip amplitudes vanish.

The assumption that production amplitudes in $\pi^- p \to  
\pi^0\pi^0 n$ do not depend on nucleon spin is in conflict with the  
general consensus that hadronic reactions depend on nucleon spin up  
to the highest energies, and contradicts all that we have learned  
from measurements of $\pi N \to \pi^- \pi^+ N$ on polarized targets  
at CERN. Applied to reactions $\pi^- p \to \pi^+ \pi^- n$ and $\pi^+  
n \to \pi^+ \pi^- p$, the assumption has observable consequences  
that can be tested directly in measurements with polarized targets.

The first consequence is that all polarized moments $p^L_M$ vanish  
identically. All experiments on polarized targets however found  
large nonzero polarized moments. An example is given in Fig.~1 which  
shows polarized target asymmetry $A$ related to the moment $p^0_0$.  
The polarized target asymmetry has large nonzero (negative) values  
in both reactions. Measurements of $K^+ n \to K^+ \pi^- p$ show  
similarly large values of $A$\cite{lesquen89}.

The experiments on polarized targets are best analysed using  
nucleon transversity amplitudes rather than nucleon helicity  
amplitudes. The second consequence of the assumption of independence  
of production amplitudes on nucleon spin is that all transversity  
amplitudes $|\overline A|$ with recoil nucleon transversity ``up''  
are equal in magnitude to transversity amplitudes $|A|$ with recoil  
nucleon transversity ``down'' relative to the scattering plane  
$\pi^- N \to (\pi^-\pi^+)N$. In Fig.~2 we show the ratios of  
transversity amplitudes for $S$-, $P$-, $D$- and $F$-waves for  
dimeson helicity $\lambda = 0$. The ratios are far from unity,  
indicating that production amplitudes depend strongly on nucleon  
spin.

If the assumption that the production amplitudes are independent of  
nucleon spin does not work in reactions $\pi^- p \to \pi^-\pi^+ n$,  
$\pi^+ n \to \pi^+ \pi^- p$ and $K^+ n \to K^+ \pi^- p$ then there  
is no reason to assume that it will work in $\pi^- p \to \pi^0\pi^0  
n$ reaction. We must conclude that some of the results of the  
analyses of $\pi^- p \to \pi^0\pi^0 n$ by GAMS and E852  
collaborations are not reliable.

The question of reliability of amplitude analyses based on  
assumption of independence of production amplitudes on nucleon spin  
is of special importance to confirmation and further study of the  
narrow $\sigma(750)$ state in $\pi^- p \to \pi^0\pi^0 n$ reaction.  
The evidence for narrow $\sigma(750)$ is closely connected to the  
spin dependence of production amplitudes. In Fig.~3 we show the two  
$S$-wave production amplitudes for $\pi^- p \to \pi^- \pi^+ n$. We  
see that while the transversity up amplitude $|\overline S|^2\Sigma$  
resonates in both solutions around 750 MeV the transversity down  
amplitude $|S|^2\Sigma$ is large and non-resonating. This results in  
a partial wave intensity $I_S = (|S|^2 + |\overline S|^2)\Sigma$  
that does not necessarily show a narrow resonant behaviour. As seen  
in Fig.~4, such is the case of solution $I_S(2,2)$.

It is therefore necessary to establish what quantities can be  
determined from the measurements of $\pi^- p \to \pi^0\pi^0 n$ on  
unpolarized targets without the assumption of independence of  
production amplitudes on nucleon spin. Furthermore, it is necessary  
to find out if a model independent amplitude analysis of $\pi^- p  
\to \pi^0\pi^0 n$ in measurements on polarized targets is possible.  
The purpose of this work is to provide answers to these questions.  
We shall show that in measurements of $\pi^- p \to \pi^0\pi^0 n$ on  
unpolarized targets in the region where $S$- and $D$-wave dominate,  
one can measure four spin-averaged partial wave intensities and  
three unrelated phases connected with the spin-averaged interference  
terms. We will also show that model independent amplitude analysis  
is possible when measurements of $\pi^- p \to \pi^0\pi^0$ are made  
on polarized target, both in the region where $S$- and $D$-wave  
dominate as well as in the region where $G$-wave also contributes.  
We shall propose that such measurements are a natural extension of  
measurements on unpolarized targets and should be performed at both  
IHEP in Protvino and at BNL using Brookhaven Multi Particle  
Spectrometer.

The paper is organized in seven sections. The kinematics,  
observables and pion production amplitude are introduced in Section  
II. The method of model independent analysis of data on unpolarized  
target is described in Section III. In Section IV we compare this  
method with model dependent analyses of GAMS and E852  
Collaborations. In Section V we describe a model-independent  
amplitude analysis of $\pi^- p \to \pi^0\pi^0 n$ on polarized target  
in the absence of $G$-wave. In Section VI we extend the  
model-independent amplitude analysis to include the $G$-wave  
amplitudes. The paper closes with a summary and proposals for  
measurements of $\pi^- p\to \pi^0\pi^0 n$ and $\pi^- p \to \eta\eta  
n$ on polarized targets in Section VII.

\section{Kinematics, observables, and amplitudes}

\subsection{Kinematics}

Various aspects of phase space, kinematics and amplitudes in pion  
production in $\pi N \to \pi\pi N$ reactions are described in  
several books\cite{pilkuhn67,bycling73,humble74}. The kinematical  
variables used to describe the dimeson production on a polarized  
target at rest are $(s,t,m,\theta,\phi,\psi,\delta)$ where $s$ is  
the c.m.s. energy squared, $t$ is the four-momentum transfer to the  
nucleon squared, and $m$ is the dimeson invariant mass. The angles  
$(\theta,\phi)$ describe the direction of $\pi^0$ in the  
$\pi^0\pi^0$ rest frame. The angle $\psi$ is the angle between the  
direction of target transverse polarization and the normal to the  
scattering plane (Fig.~5). The angle $\delta$ is the angle between  
the direction of target polarization vector and its transverse  
component (projection of polarization vector into the $x,y$ plane).  
The analysis is usually carried out in the $t$-channel helicity  
frame for the $\pi^0\pi^0$ dimeson system. The helicities of the  
initial and final nucleons are always defined in the $s$-channel  
helicity frame.

\subsection{Observables}

In our discussion of observables measured in $\pi^- p \to  
\pi^0\pi^0 n$ with polarized targets we follow the notation of Lutz  
and Rybicki\cite{lutz78}. When the polarization of the recoil  
nucleon is not measured, the unnormalized angular distribution  
$I(\theta,\phi,\psi,\delta)$ of $\pi^0\pi^0$ (or $\eta\eta$)  
production on polarized nucleons at rest of fixed $s$, $m$ and $t$  
can be written as

\begin{equation}
I(\Omega,\psi,\delta) = I_U (\Omega) + P_T \cos\psi I_C (\Omega) +  
P_T \sin\psi I_S (\Omega) + P_L I_L (\Omega)
\end{equation}

\noindent
where $P_T = P\cos\delta$ and $P_L = P \sin\delta$ are the  
transverse and longitudinal components of target polarization ${\vec  
P}$ with respect to the incident momentum (Fig.~5). The simple  
$\cos\psi$ and $\sin\psi$ dependence is due to spin ${1\over 2}$ of  
the target nucleon\cite{lutz78,bourrely80}. Parity conservation  
requires $I_U$ and $I_C$ to be symmetric, and $I_S$ and $I_L$ to be  
antisymmetric in $\phi$. In the data analysis of angular  
distribution of the dimeson system, it is convenient to use  
expansions of the angular distributions into spherical harmonics. In  
the notation of Lutz and Rybicki we have

\begin{equation}
I_U(\Omega) = \sum\limits_{L,M} t^L_M {\rm Re} Y^L_M (\Omega)
\end{equation}

\[
I_C (\Omega) = \sum\limits_{L,M} p^L_M {\rm Re} Y^L_M (\Omega)
\]

\[
I_S (\Omega) = \sum\limits_{L,M} r^L_M {\rm Im} Y^L_M (\Omega)
\]

\[
I_L (\Omega) = \sum\limits_{L,M} q^L_M {\rm Im} Y^L_M (\Omega)]
\]

\noindent
The expansion coefficients $t,p,r,q$ are called multipole moments.  
The moments $t^L_M$ are unpolarized. The moments $p^L_M$, $r^L_M$  
and $q^L_M$ are polarized moments. Experiments with transversely  
polarized targets measure only transverse moments $p^L_M$ and  
$r^L_M$ but not the longitudinal moments $q^L_M$.

The multipole moments are obtained from the experimentally observed  
distributions in each $(m,t)$ bin by means of optimization of  
maximum likelihood function which takes into account the acceptance  
of the apparatus\cite{grayer74,eadie71}. In these fits it is usually  
assumed that moments with $M>2$ vanish. However, it was pointed out  
by Sakrejda\cite{sakrejda84} that moments up to $M=4$ may have to  
be taken into account at larger momentum transfers extending to 1.0  
(GeV/c)$^2$.

The expansion coefficients $t,p,r,q$ are simply connected to  
moments of angular distributions\cite{lutz78}:

\begin{equation}
t^L_M = \epsilon_M <{\rm Re}^L_M> = {{\epsilon_M}\over{2\pi}} \int  
I (\Omega,\psi,\delta) {\rm Re} Y^L_M (\Omega) d\Omega^\prime
\end{equation}

\[
p^L_M = 2\epsilon_M <\cos\psi {\rm Re}^L_M> =  
{{2\epsilon_M}\over{2\pi}} \int I (\Omega,\psi,\delta) {\rm Re}  
Y^L_M (\Omega) \cos\psi \cos\delta d \Omega^\prime
\]

\[
r^L_M = 4 < \sin\psi {\rm Im} Y^L_M> = {4\over{2\pi}} \int I  
(\Omega,\psi,\delta) {\rm Im} Y^L_M \sin\psi \cos\delta  
d\Omega^\prime
\]

\[
q^L_M = 4 <{\rm Im} Y^L_M>= {4\over{2\pi}} \int I  
(\Omega,\psi,\delta) {\rm Im} Y^L_M \sin\delta d\Omega^\prime
\]

\noindent
where $d\Omega^\prime = d\Omega d\psi d(-\sin\delta)$. In (2.3),  
$\epsilon_M = 1$ for $M=0$ and $\epsilon_M = 2$ for $M\ne 0$.  
Integrated over the solid angles $(\theta,\phi)$, the distribution  
(2.1) becomes

\begin{equation}
I(\psi,\delta) = (1 + A P_T \cos\psi) {{d^2\sigma}\over{dmdt}}
\end{equation}

\noindent
where $A = A(s,t,m) = \sqrt{4\pi} p^0_0$ is the polarized target  
asymmetry analogous to the polarization parameter measured in  
two-body reactions. In (2.4) $d^2\sigma/dmdt$ is the integrated  
reaction cross-section

\begin{equation}
{{d^2\sigma (s,t,m)}\over{dmdt}} = \int I (\Omega,\psi,\delta)  
d\Omega^\prime
\end{equation}

\noindent
Finally we note the relation of moments $t^L_M$ to moments $H(LM)$  
introduced by Chung\cite{chung96,chung96b}:

\begin{equation}
t^L_M = \epsilon_M <{\rm Re} Y^L_M> = \epsilon_M  
\sqrt{{{2L+1}\over{4\pi}}} H (LM)
\end{equation}

\subsection{Amplitudes}

The reaction $\pi^- p\to \pi^0\pi^0 n$ is described by production  
amplitudes $H_{\lambda_n, 0\lambda_p} (s,t,m,\theta,\phi)$ where  
$\lambda_p$ and $\lambda_n$ are the helicities of the proton and  
neutron, respectively. The production amplitudes can be expressed in  
terms of production amplitudes corresponding to definite dimeson  
spin $J$ using an angular expansion

\begin{equation}
H_{\lambda_n, 0\lambda_p} = \sum\limits^\infty_{J=0}  
\sum\limits^{+J}_{\lambda = -J} (2J + 1)^{1/2}  
H^J_{\lambda\lambda_n, 0\lambda_p} (s,t,m) d^J_{\lambda 0} (\theta)  
e^{i\lambda\phi}
\end{equation}

\noindent
where $J$ is the spin and $\lambda$ the helicity of the  
$(\pi^0\pi^0)$ dimeson system. Because of the identity of the two  
final-state mesons, the ``partial waves'' with odd $J$ are absent so  
that $J=0,2,4,\ldots$

In the following we will consider only $S$-wave $(J=0)$, $D$-wave  
$(J=2)$ and $G$-wave $(J=4)$ amplitudes. Furthermore, we will  
restrict the dimeson helicity $\lambda$ to values $\lambda=0$ or  
$\pm 1$ only in accordance with the assumption that moments with  
$M>2$ vanish. This assumption is supported by experiments.

The ``partial wave'' amplitudes $H^J_{\lambda\lambda_n,0\lambda_p}$  
can be expressed in terms of nucleon helicity amplitudes with  
definite $t$-channel exchange naturality. The nucleon $s$-channel  
helicity amplitudes describing the production of $(\pi^0\pi^0)$ (or  
$(\eta\eta)$) system in the $S$-, $D$- and $G$-wave states are:

\begin{equation}
0^- {1\over 2}^+ \to 0^+ {1\over 2}^+\ \colon\ H^0_{0+,0+} = S_0,\  
H^0_{0+,0-} = S_1
\end{equation}

\[
0^- {1\over 2}^+ \to 2^+ {1\over 2}^+\ \colon\ H^2_{0+,0+} =  
D^0_0,\ H^2_{0+,0-} = D^0_1
\]

\[
H^2_{\pm 1+,0+} = {{D^+_0 \pm D^-_0}\over{\sqrt 2}},\ H^2_{\pm 1+,  
0-} = {{D^+_1 \pm D^-_1}\over{\sqrt 2}}
\]

\[
0^- {1\over 2}^+ \to 4^+ {1\over 2}^+\ \colon\ H^4_{0+,0+} =  
G^0_0,\ H^4_{0+,0-} = G^0_1
\]

\[
H^4_{\pm 1 +, 0+} = {{G^+_0 \pm G^-_0}\over{\sqrt 2}},\ H^4_{\pm  
1+,0-} = {{G^+_1 \pm G^-_1}\over{\sqrt 2}}
\]

\noindent
At large $s$, the amplitudes $S_n$, $D^0_n$, $D^-_n$, $G^0_n$,  
$G^-_n$, $n=0,1$ are dominated by unnatural exchanges while the  
amplitudes $D^+_n$ and $G^+_n$, $n=0,1$ are dominated by natural  
exchanges. The index $n = |\lambda_p - \lambda_n|$ is nucleon  
helicity flip.

The observables obtained in experiments on transversely polarized  
targets in which recoil nucleon polarization is not observed are  
most simply related to nucleon transversity amplitudes of definite  
naturality\cite{lutz78,lesquen89,kotanski66}. For $S$-, $D$- and  
$G$-waves they are defined as follows:

\begin{equation}
S = k(S_0 + i S_1)\ ,\ \overline S = k(S_0 - i S_1)
\end{equation}

\[
D^0 = k (D^0_0 + iD^0_1)\ , \ \overline D^0 = k (D^0_0 - i D^0_1)
\]

\[
D^- = k (D^-_0 + iD^-_1)\ ,\ \overline D^- = k (D^-_0 - iD^-_1)
\]

\[
D^+ = k(D^+_0 - i D^+_1)\ ,\ \overline D^+ = k(D^+_0 + iD^+_1)
\]

\[
G^0 = k(G^0_0 + i G^0_1)\ ,\ \overline G^0 = k(G^0_0 - iG^0_1)
\]

\[
G^- = k (G^-_0 + iG^-_1)\ ,\ \overline G^- = k(G^-_0 - iG^-_1)
\]

\[
G^+ = k (G^+_0 - iG^+_1)\ ,\ \overline G^+ = k(G^+_0 + iG^+_1)
\]

\noindent
where $k=1/\sqrt 2$. The formal proof that the amplitudes defined  
in (2.9) are actually transversity amplitudes is given from  
definition in the Appendix in Ref.~19.

The nucleon helicity and nucleon transversity amplitudes differ in  
the quantization axis for the nucleon spin. The transversity  
amplitudes $S$, $D^0$, $D^-$, $D^+$, $G^0$, $G^-$, $G^+$ $(\overline  
S, \overline D^0, \overline D^-, \overline D^+, \overline G^0,  
\overline G^-, \overline G^+)$ describe the production of the  
dimeson state with the recoil nucleon spin antiparallel or ``down''  
(parallel or ``up'') relative to the normal ${\vec n}$ to the  
production plane. The direction of normal ${\vec n}$ is defined  
according to Basel convention by ${\vec p}_\pi \times {\vec  
p}_{\pi\pi}$ where ${\vec p}_\pi$ and ${\vec p}_{\pi\pi}$ are the  
incident pion and dimeson momenta in the target proton rest frame.

Using the symbols $\uparrow$ and $\downarrow$ for the nucleon  
transversities up and down, respectively, the following table shows  
the spin states of target protons and recoil neutrons and the  
dimeson helicities corresponding to the transversity amplitudes  
(2.9):

\narrowtext
\begin{quasitable}
\begin{tabular}{lccc}
&$p$ &$n$ &$(\pi^0\pi^0)$ \\
\tableline
$S, D^0, G^0$ & $\uparrow$ &$\downarrow$ &0\\
$\overline S, \overline D^0, \overline G^0$ &$\downarrow$  
&$\uparrow$ &0\\
$D^-, G^-$ &$\uparrow$ &$\downarrow$ &$+1$ or $-1$\\
$\overline D^-, \overline G^-$ &$\downarrow$ &$\uparrow$ &$+1$ or $-1$\\
$D^+, G^+$ &$\downarrow$ &$\downarrow$ &$+1$ or $-1$\\
$\overline G^+, \overline G^+$ &$\uparrow$ &$\uparrow$ &$+1$ or $-1$
\end{tabular}
\end{quasitable}

\noindent
Parity conservation requires that in the transversity frame the  
dimeson production with helicities $\pm 1$ depends only on the  
transversities of the initial and final nucleons. The amplitudes  
$D^-$, $\overline D^-$, $\ldots$, $G^+$, $\overline G^+$ do not  
distinguish between dimeson helicity states with $\lambda = +1$ or  
$-1$. Also, the dimeson production with helicity $\lambda=0$ is  
forbidden by parity conservation when the initial and final nucleons  
have the same transversities.

\subsection{Observables in terms of amplitudes}

It is possible to express the moments $t^L_M$ and $p^L_M$ in terms  
of quantities that do not depend explicitly on whether we use  
nucleon helicity or nucleon transversity amplitudes. However,  
eventually we are going to work with transversity amplitudes. The  
quantities we shall need are spin-averaged partial wave intensity

\begin{equation}
I_A = |A|^2 +  |\overline A|^2 = |A_0|^2 + |A_1|^2
\end{equation}

\noindent
and partial wave polarization

\begin{equation}
P_A = |A|^2 - |\overline A|^2 = 2\epsilon_A {\rm Im} (A_0 A^*_1)
\end{equation}

\noindent
where $\epsilon_A = +1$ for $A=S, D^0, D^-, G^0, G^-$ and  
$\epsilon_A = -1$ for $A = D^+, G^+$. We also introduce  
spin-averaged interference terms

\begin{equation}
R(AB) = {\rm Re} (AB^* + \overline A\; \overline B^*) = {\rm Re}  
(A_0 B_0 + \epsilon_A \epsilon_B A_1 B^*_1)
\end{equation}

\begin{equation}
Q(AB) = {\rm Re} (AB^* - \overline A\; \overline B^*) = {\rm Re}  
(\epsilon_B A_0 B^*_1 - \epsilon_A A_1 B^*_0)
\end{equation}

\noindent
Then moments $t^L_M$ can be expressed in terms of spin-averaged  
intensities $I_A$ and spin-averaged interference terms $R(AB)$. The  
moments $p^L_M$ are then expressed in terms of polarizations $P_A$  
and interference terms $Q(AB)$. The formulas for $p^L_M$ are  
obtained from those for $t^L_M$ using a replacement $I_A \to  
\epsilon_A P_A$ and $R(AB)\to Q(AB)$ for $\epsilon_A = \epsilon_B =  
1$ and $R(AB)\to -Q(AB)$ for $\epsilon_A = \epsilon_B = -1$. There  
is no mixing of natural and unnatural exchange amplitudes in the  
moments $t^L_M$ and $p^L_M$.

Using the results of Lutz and Rybicki\cite{lutz78} and of  
Chung\cite{chung96b}, we obtain the following expressions for  
moments in terms quantities (2.10)--(2.13) and a constant $c  
=\sqrt{4\pi}$:

\FL
\begin{equation}
{\rm Unpolarized\ moments}
\end{equation}

\[
ct^0_0 = I_S + I_{D^0} + I_{D^-} + I_{D^+} + I_{G^0} + I_{G^-} + I_{G^+}
\]

\[
ct^2_0 = \sqrt 5 \{ {2\over{\sqrt 5}} R(SD^0) + {2\over 7} I_{D^0}  
+ {1\over 7} (I_{D^-} + I_{D^+})
\]

\[
+ {{12}\over{7\sqrt 5}} R(D^0 G^0) + {{2\sqrt 6}\over 7} [R(D^-G^-)  
+ R(D^+ G^+)]
\]

\[
+{{20}\over{77}} I_{G^0} + {{17}\over{77}} (I_{G^-} + I_{G^+})\}
\]

\[
ct^2_1 = 2\sqrt 5 \{ {2\over{\sqrt{10}}} R(SD^-) + {{\sqrt  
2}\over{7}} R(D^0 D^-) + {{2\sqrt 3}\over{7}} R(D^0 G^-)
\]

\[
-{4\over 5} \sqrt{{2\over 5}} R(D^- G^0) + {{2\sqrt{15}}\over{77}}  
R(G^0 G^-)\}
\]

\[
ct^2_2 = 2\sqrt 5 \{ {1\over 7} \sqrt{{3\over 2}} (I_{D^-} -  
I_{D^+}) - {1\over 7} [R(D^- G^-) - R(D^+ G^+)]
\]

\[
+ {{5\sqrt 6}\over{77}} (I_{G^-} - I_{G^+})\}
\]

\[
ct^4_0 = \sqrt 9 \{ {2\over 7} I_{D^0} - {4\over{21}} (I_{D^-} +  
I_{D^+}) + {2\over 3} R(SG^0)+
\]

\[
+ {{40\sqrt 5}\over{231}} R(D^0 G^0) + {{162}\over{1001}} I_{G^0} +
\]

\[
+ {{10}\over{77}} \sqrt{{2\over 3}} [R(D^- G^-) + R(D^+ G^+)] +  
{{81}\over{1001}} (I_{G^-} + I_{G^+})\}
\]

\[
ct^4_1 = 2\sqrt 9 \{ {2\over 7} \sqrt{{5\over 3}} R(D^0 D^-) +  
{{\sqrt 2}\over 3} R(SG^-) + {{17\sqrt{10}}\over{231}} R(D^0 G^-)
\]

\[
+ {{10}\over{77\sqrt 3}} R(D^- G^0) + {{81\sqrt 2}\over{1001}}  
R(G^0 G^-)\}
\]

\[
ct^4_2 = 2\sqrt 9 \{ {{\sqrt{10}}\over{21}} (I_{D^-} - I_{D^+}) +  
{{6\sqrt{15}}\over{154}} [R(D^- G^-) - R(D^+ G^+)]
\]

\[
+ {{27\sqrt{10}}\over{1001}} (I_{G^-} - I_{G^+})\}
\]

\[
ct^6_0 = \sqrt{13} \{ {{30\sqrt 5}\over{143}} R(D^0 G^0) -  
{{20\sqrt 6}\over{143}} [R(D^- G^-) + R(D^+ G^+)]
\]

\[
+ {{20}\over{143}} I_{G^0} - {1\over{143}} (I_{G^-} + I_{G^+})\}
\]

\[
ct^6_1 = 2\sqrt{13} \{ {{10\sqrt{21}}\over{143}} R(D^0 G^-) +  
{{10\sqrt{35}}\over{143\sqrt 2}} R(D^- G^0)+  
{{2\sqrt{105}}\over{143}} R(D^0 G^-)\}
\]

\[
ct^6_2 = 2\sqrt{13} \{ {{4\sqrt{70}}\over{143}} [R(D^- G^-) - R(D^+  
G^+)]+ {{\sqrt{105}}\over{143}} (I_{G^-} - I_{G^+})\}
\]

\[
ct^8_0 = \sqrt{17} \{ {{490}\over{2431}} I_{G^0} -  
{{392}\over{2431}} (I_{G^-} + I_{G^+})\}
\]

\[
ct^8_1 = 2\sqrt{17} \{ {{294\sqrt 5}\over{2431}} R(G^0 G^-)\}
\]

\[
ct^8_2 = 2\sqrt{17} \{ {{42\sqrt{35}}\over{2431}} (I_{G^-} - I_{G^+})\}
\]

\FL
\begin{equation}
{\rm Polarized\ moments}\  p^L_M
\end{equation}

\[
cp^0_0 = P_S + P_{D^0} + P_{D^-} - P_{D^+} + P_{G^0} + P_{G^-} - P_{G^+}
\]

\[
cp^2_0 = \sqrt 5 \{ {2\over{\sqrt 5}} Q(SD^0) + {2\over 7} P_{D^0}  
+ {1\over 7} (P_{D^-} - P_{D^+}) +
\]

\[
+ {{12}\over{7\sqrt 5}} Q(D^0 G^0) + {{2\sqrt 6}\over{7}}  
[Q(D^-G^-) - Q(D^+ G^+)] +
\]

\[
+ {{20}\over{77}} P_{G^0} + {{17}\over{77}} (P_{G^-} - P_{G^+})\}
\]

\[
cp^2_1 = 2\sqrt 5 \{ {2\over{\sqrt{10}}} Q(SD^-) + {{\sqrt 2}\over  
7} Q(D^0 D^-) + {{2\sqrt 3}\over 7} Q(D^0 G^-)
\]

\[
-{4\over 5} \sqrt{{2\over 5}} Q (D^- G^0) + {{2\sqrt{15}}\over{77}}  
Q(G^0 G^-)\}
\]

\[
cp^2_2 = 2\sqrt 5 \{ {1\over 7} \sqrt{{3\over 2}} (P_{D^-} +  
P_{D^+} ) - {1\over 7} [Q(D^- G^-) + Q(D^+ G^+)]
\]

\[
+ {{5\sqrt 6}\over{77}} (P_{G^-} + P_{G^+})\}
\]

\[
cp^4_0 = \sqrt 9 \{ {2\over 7} P_{D^0} - {4\over{21}} (P_{D^-} -  
P_{D^+} ) + {2\over 3} Q(SG^0)
\]

\[
+ {{40\sqrt 5}\over{231}} Q(D^0 G^0) + {{162}\over{1001}} P_{G^0} +
\]

\[
+ {{10}\over{77}} \sqrt{{2\over 3}} [Q(D^- G^-) - Q(D^+ G^+)] +  
{{81}\over{1001}} (P_{G^-} - P_{G^+})\}
\]

\[
cp^4_1 = 2\sqrt 9 \{ {2\over 7} \sqrt{{5\over 3}} Q(D^0 D^-) +  
{{\sqrt 2}\over 3} Q(SG^-) + {{17\sqrt{10}}\over{231}} Q(D^0 G^-)
\]

\[
+ {{10}\over{77\sqrt 3}} Q(D^- G^0) + {{81\sqrt 2}\over{1001}}  
Q(G^0 G^-)\}
\]

\[
cp^4_2 = 2\sqrt 9 \{ {{\sqrt{10}}\over{21}} (P_{D^-} + P_{D^+}) +  
{{6\sqrt{15}}\over{154}} [Q(D^-G^-) + Q(D^+ G^+)]
\]

\[
+ {{27\sqrt{10}}\over{1001}} (P_{G^-} + P_{G^+})\}
\]

\[
cp^6_0 = \sqrt{13} \{ {{30\sqrt 5}\over{143}} Q(D^0 G^0) -  
{{20\sqrt 6}\over{143}} [Q(D^- G^-) - Q(D^+ G^+)]
\]

\[
+ {{20}\over{143}} P_{G^0} - {1\over{143}} (P_{G^-} - P_{G^+} )\}
\]

\[
cp^6_1 = 2\sqrt{13} \{ {{10\sqrt{21}}\over{143}} Q(D^0G^-) +  
{{10\sqrt{35}}\over{143\sqrt 2}} Q(D^- G^0)+  
{{2\sqrt{105}}\over{143}} Q(G^0 G^-)\}
\]

\[
cp^6_2 = 2\sqrt{13} \{ {{4\sqrt{70}}\over{143}} [Q(D^- G^-) + Q(D^+  
G^+)] + {{\sqrt{105}}\over{143}} (P_{G^-} + P_{G^+})\}
\]

\[
cp^8_0 = \sqrt{17} \{ {{490}\over{2431}} P_{G^0} -  
{{392}\over{2431}} (P_{G^-} - P_{G^+})\}
\]

\[
cp^8_1 = 2\sqrt{17} \{ {{294\sqrt 5}\over{2431}} Q(G^0 G^-)\}
\]

\[
cp^8_2 = 2\sqrt{17} \{ {{42\sqrt{35}}\over{2431}} (P_{G^-} + P_{G^+})\}
\]

\FL
\begin{equation}
{\rm Polarized\  moments}\  r^L_M
\end{equation}

\[
r^2_1 = 2\sqrt 2 {\rm Re} (SD^{+*} - \overline S\; \overline D^{+*}  
) + {{2\sqrt{10}}\over 7} {\rm Re} (D^0 D^{+*} - \overline D^0  
\overline D^{+*})
\]

\[
r^2_2 = {{2\sqrt{30}}\over{7}} {\rm Re} (D^- D^{+*} - \overline D^-  
\overline D^{+*})
\]

\[
r^4_1 = - {{4\sqrt{15}}\over{7}} {\rm Re} (D^0 D^{+*} - \overline  
D^0 \overline D^{+*})
\]

\[
r^4_2 = - {{4\sqrt{10}}\over{7}} {\rm Re} (D^- D^{+*} - \overline  
D^- \overline D^{+*})
\]

We do not include $G$-wave contributions in the polarized moments  
$r^L_M$. In general, these moments are not well determined in  
measurements on transversely polarized targets and, as can be seen  
in Appendix A, the calculation of relative phases between the  
natural exchange amplitude $D^+$ and the unnatural exchange  
amplitudes $S, D^0, D^-$ already involves high degree of ambiguity.  
The inclusion of $G$ waves would make the situation even less  
tractable.

\section{Model independent analysis of measurements on unpolarized  
targets.}

We will now show that in the mass region where only $S$- and  
$D$-waves dominate, {\sl i.e.,} up to about 1500 MeV, it is possible  
to perform an analysis of measurements of $\pi^- p \to \pi^0\pi^0  
n$ and $\pi^- p \to \eta\eta n$ on unpolarized targets without the  
simplifying assumption that production amplitudes do not depend on  
nucleon spin. However we will find that data on unpolarized targets  
measure in a model independent way only the partial wave intensities  
and three unrelated interference phases, and not the production  
amplitudes which remain undetermined.

When only $S$- and $D$-wave contribute, the unpolarized moments are  
(with $c=\sqrt{4\pi}$):

\begin{equation}
ct^0_0 = I_S + I_{D^0} + I_{D^-} + I_{D^+}
\end{equation}

\[
ct^2_0 = 2R(SD^0) + {{2\sqrt 5}\over 7} I_{D^0} + {{\sqrt 5}\over  
7} (I_{D^-} + I_{D^+})
\]

\[
ct^2_1 =  2\sqrt 2 R(SD^-) + {{2\sqrt{10}}\over 7} R(D^0 D^-)
\]

\[
ct^2_2 = {{\sqrt{30}}\over 7} (I_{D^-} - I_{D^+})
\]

\[
ct^4_0 = {6\over 7} I_{D^0} - {4\over 7} (I_{D^-} + I_{D^+})
\]

\[
ct^4_1 = {{4\sqrt{15}}\over 7} R(D^0 D^-)
\]

\[
ct^4_2 = {{2\sqrt{10}}\over 7} (I_{D^-} - I_{D^+})
\]

\noindent
There are 6 independent observables to determine 7 unknowns -- 4  
partial wave intensities and 3 spin averaged interference terms.  
Since there are more unknowns than observables, it is necessary to  
express the maximum likelihood function ${\cal L}$ in terms of the  
partial wave intensities and the interference terms and fit ${\cal  
L}$ to observed data to find a solution.

For this purpose we will now show that the interference terms  
$R(AB)$ in (3.1) have a general form

\begin{equation}
R(AB) = \sqrt I_A \sqrt I_B \cos (\delta_{AB})
\end{equation}

\noindent
>From the definition (2.12) we have

\[
R(AB) = \sum\limits^1_{n=0} {\rm Re} (A_n B^*_n) =  
\sum\limits^1_{n=0} |A_n| |B_n| \cos (\phi^A_n - \phi^B_n)
\]

\noindent
We can write

\begin{equation}
R(AB) = \sqrt I_A \sqrt I_B Z_{AB}
\end{equation}

\noindent
With definitions for $n=0,1$

\begin{equation}
\xi^{AB}_n = {{|A_n|}\over{\sqrt I_A}} {{|B_n|}\over{\sqrt I_B}}\  
,\ \varphi^{AB}_n = \phi^A_n - \phi^B_n
\end{equation}

\noindent
we have

\begin{equation}
Z_{AB} = \xi^{AB}_0 \cos \varphi^{AB}_0 + \xi^{AB}_1 \cos \varphi^{AB}_1
\end{equation}

\noindent
We now recall a theorem from wave theory\cite{bronshtein54}

\begin{equation}
A_1 \sin (\omega t + \varphi_1) + A_2 \sin (\omega t + \varphi_2) =  
A \sin (\omega t + \varphi)
\end{equation}

\noindent
where

\begin{equation}
A^2 = A^2_1 + A^2_2 + 2A_1 A_2 \cos (\varphi_2 - \varphi_1)
\end{equation}

\[
\tan\varphi = {{A_1 \sin \varphi_1 + A_2\sin \varphi_2}\over{A_1  
\cos \varphi_1 + A_2 \cos \varphi_2}}
\]

\noindent
For $\omega t = {\pi\over 2}$ we get

\begin{equation}
A_1 \cos \varphi_1 + A_2 \cos \varphi_2 = A \cos \varphi
\end{equation}

\noindent
with $A$ and $\varphi$ given above. We can apply (3.8) to (3.5) and get

\[
Z_{AB} = \xi_{AB} \cos \delta_{AB}
\]

\noindent
where $\xi_{AB}$ and $\delta_{AB}$ are given by (3.7) with the  
appropriate substitutions from (3.5). After some algebra it is  
possible to show that

\begin{equation}
0\le \xi_{AB} \le + 1
\end{equation}

\noindent
so that $-1 \le Z_{AB} \le +1$. Thus we can actually write $Z_{AB}  
\equiv \cos\delta_{AB}$ which proves the statement (3.2). The phase  
$\delta_{AB}$ is not simply related to the two relative phases  
$\phi^A_0 - \phi^B_0$ and $\phi^A_1 - \phi^B_1$ of the helicity  
amplitudes $A_n, B_n, n=0,1$. Moreover, $\cos\delta_{AB}$ is a  
measurable parameter along with the intensities $I_A$ and $I_B$.

We will refer to $\delta_{SD^0}$, $\delta_{SD^-}$ and  
$\delta_{D^0D^-}$ in (3.1) as interference phases. Notice again that  
interference phases are not relative phases between amplitudes and  
are thus independent. Whereas relative phases satisfy for $n=0,1$

\begin{equation}
(\phi^S_n - \phi^{D^0}_n) + (\phi^{D^-}_n - \phi^S_n) +  
(\phi^{D^0}_n - \phi^{D^-}_n) = 0
\end{equation}

\noindent
there is no such relation for the interference phases.

We can use (3.2) to express the maximum likelihood function ${\cal  
L}$ in terms of the 4 intensities $I_S, I_{D^0}, I_{D^-}, I_{D^+}$  
and 3 interference phases $\delta_{SD^0}$, $\delta_{SD^-}$ and  
$\delta_{D^0D^-}$ and fit ${\cal L}$ to the observed angular  
distributions to find a solution for these quantities in each  
$(m,t)$ bin. We can conclude that analysis of data on $\pi^- p \to  
\pi^0\pi^0 n$ unpolarized target is possible without the assumption  
that production amplitudes are independent of nucleon spin. However  
the data on unpolarized target cannot determine the 8 moduli and 6  
cosines of dependent relative phases of production amplitudes. As we  
show below, for that determination measurements on polarized target  
are necessary. The measurements on unpolarized target determine  
only 4 partial wave intensities and 3 interference phases in a model  
independent way.

In a mass region where $G$-waves contribute, measurements on  
unpolarized target measure 12 independent unpolarized moments  
$t^L_M$. There are 7 intensities and 11 spin averaged interference  
terms in (2.14) for a total of 18 unknowns. In this case model  
independent amplitude analysis is not possible. However we shall see  
below that model independent analysis including $G$-waves is  
possible for measurements on polarized targets.

\section{Comparison with model dependent analyses of $\pi^-  
\lowercase{p} \to \pi^0\pi^0 \lowercase{n}$ on unpolarized target.}

Both GAMS Collaboration and BNL E852 Collaboration use the  
assumption of independence of production amplitudes on nucleon  
spin\cite{chung96,chung96b} but employ different strategies in  
actual fits to the observed angular  
distributions\cite{kondashov,gunter}. We will confine our discussion  
to the mass region where $S$- and $D$-waves dominate.

The assumption of independence of production amplitudes on nucleon  
spin means that formally there is one $S$-wave amplitude $S$ and  
three $D$-wave amplitudes $D^0$, $D^-$, $D^+$. The amplitudes have  
no nucleon spin index. However, as we have argued above, these  
amplitudes are essentially the single flip helicity amplitudes  
$(n=1)$ while all helicity non-flip amplitudes $(n=0)$ are assumed  
to vanish.

In the GAMS approach\cite{kondashov} the unpolarized moments are  
then written as (with $c=\sqrt{4\pi}$)

\begin{equation}
ct^0_0 = |S|^2 + |D^0|^2 + |D^-|^2 + |D^+|^2
\end{equation}

\[
ct^2_0 = 2 {\rm Re} (SD^{0*}) + {{2\sqrt 5}\over 7} |D^0|^2 +  
{{\sqrt 5}\over 7} (|D^-|^2 + |D^+|^2)
\]

\[
ct^2_1 = 2\sqrt 2 {\rm Re} (SD^{-*}) + {{2\sqrt{10}}\over 7} {\rm  
Re} (D^0 D^{-*})
\]

\[
ct^2_2 = {{\sqrt{30}}\over 7} (|D^-|^2 - |D^+|^2)
\]

\[
ct^4_0 = {6\over 7} |D^0|^2 - {4\over 7} (|D^-|^2 + |D^+|^2)
\]

\[
ct^4_1 = {4\over 7} \sqrt{15} {\rm Re} (D^0 D^{-*})
\]

\[
ct^4_2 = {2\over 7} \sqrt{10} (|D^-|^2 - |D^+|^2)
\]

There are 6 independent equations for 7 unknowns -- 4 moduli and 3  
cosines of relative phases. The GAMS Collaboration determines these  
quantities by expressing the maximum likelihood function ${\cal L}$  
in terms of the amplitudes (moduli and cosines) and fitting ${\cal  
L}$ to the observed angular distribution to find solutions for the  
moduli and relative phases\cite{kondashov95,kondashov96,kondashov}.  
Formally this approach is equivalent to our approach (described in  
the previous Section) with an additional assumption that the  
interference phases are not independent but satisfy a constraint

\begin{equation}
\delta_{SD^0} + \delta_{D^-S} + \delta_{D^0D^-} = 0
\end{equation}

\noindent
What GAMS Collaboration is actually doing is determining partial  
wave intensities $I_A, A=S_1$, $D^0, D^-, D^+$ and interference  
phases subject to the constraint (4.2). When the constraint (4.2) is  
removed, their approach becomes fully model independent  
determination but not of amplitudes but of partial wave intensities.

The BNL E852 employs a different approach\cite{gunter}. They  
express the moduli squared and interference terms in (4.1) in terms  
of real and imaginary parts for amplitudes $S, D^0$ and $D^-$. Since  
there is no interference with $D^+$, only $|D^+|^2$ is retained.  
Thus there are 7 unknown quantities. The maximum likelihood function  
is then expressed in terms of these unknown real and imaginary  
parts of $S, D^0, D^-$ and $|D^+|^2$ and fitted to the observed  
angular distributions to find the solution for the  
amplitudes\cite{gunter}. Formally this approach is different from  
our model independent method and relies more explicitly on the  
assumption that the non-flip helicity amplitudes all vanish.

\section{Model-independent amplitude analysis of $\pi^-  
\lowercase{p}_\uparrow \to \pi^0\pi^0 \lowercase{n}$ measured on  
polarized target with $G$-wave absent.}

In the following we will assume that unpolarized and polarized  
moments $t^L_M$ and $p^L_M$ (and $r^L_M$) have been determined using  
maximum likelihood method in data analysis of measurements of  
$\pi^- p \to \pi^0\pi^0 n$ and $\pi p \to \eta\eta n$ on polarized  
targets in a manner previously used in reactions $\pi N_\uparrow \to  
\pi^-\pi^+  
N$\cite{grayer74,groot78,becker79,becker79b,chabaud83,sakrejda84,rybicki85,lesquen85,lesquen89}.  
In this Section we show that analytical solution exists for $S$ and  
$D$ wave in mass region where these waves dominate. In the next  
section we extend the solution to include the $G$-wave amplitudes.  
In both cases we will find it useful to work with nucleon  
transversity amplitudes (2.9).

In the mass region where $S$- and $D$-waves dominate and the  
$G$-wave is absent, there are 7 unpolarized moments $t^L_M$, 7  
polarized moments $p^L_M$ and 4 polarized moments $r^L_M$ measured  
in each $(m,t)$ bin. Looking at equations (2.14) and (2.15), and  
recalling definitions (2.10)--(2.13), we see that it is advantageous  
to introduce new observables which are the sum and the difference  
of corresponding moments $t^L_M$ and $p^L_M$. We thus define (with  
$c=\sqrt{4\pi}$) the first set of equations

\begin{equation}
a_1 = {c\over 2} (t^0_0 + p^0_0) = |S|^2 + |D^0|^2 + |D^-|^2 +  
|\overline D^+|^2
\end{equation}

\[
a_2 = {c\over 2} (t^2_0 + p^2_0) = 2 {\rm Re} (SD^{0*}) + {{2\sqrt  
5}\over 7} |D^0|^2 + {{\sqrt 5}\over 7} (|D^-|^2 + |D^+|)
\]

\[
a_3 = {c\over 2} (t^2_1 + p^2_1) = 2\sqrt 2 {\rm Re} (SD^{-*}) +  
{{2\sqrt{10}}\over 7} {\rm Re} (D^0 D^{-*})
\]

\[
a_4 = {c\over 2} (t^2_2 + p^2_2) = {{\sqrt{30}}\over 7} (|D^-|^2 -  
|\overline D^+|^2)
\]

\[
a_5 = {c\over 2} (t^4_0 + p^4_0) = {6\over 7} |D^0|^2 - {4\over 7}  
(|D^-|^2 + |D^+|^2)
\]

\[
a_6 = {c\over 2} (t^4_1 + p^4_1) = {4\over 7} \sqrt{15} {\rm Re}  
(D^0 D^{-*})
\]

\[
a_7 = {c\over 2} (t^4_2 + p^4_2) = {2\over 7} \sqrt{10} (|D^-|^2 -  
|\overline D^+|^2)
\]

\noindent
The second set of equations is obtained by defining observables  
$\overline a_1, \overline a_2, \ldots ,\overline a_7$ which are the  
difference of corresponding moments. We obtain

\begin{equation}
\overline a_1 = {c\over 2} (t^0_0 - p^0_0) = |\overline S|^2 +  
|\overline D^0|^2 + |\overline D^-|^2 + |D^+|^2
\end{equation}

\[
\overline a_2 = {c\over 2} (t^2_0 - p^2_0) = 2 {\rm Re} (\overline  
S\; \overline D^{0*}) + {{2\sqrt 5}\over 7} |\overline D^0|^2 +  
{{\sqrt 5}\over 7} (|\overline D^-|^2 + |D^+|^2)
\]

\[
\overline a_3 = {c\over 2} (t^2_1 - p^2_1) = 2\sqrt 2 {\rm Re}  
(\overline S\;  \overline D^{-*}) + {{2\sqrt{10}}\over 7} {\rm Re}  
(\overline D^0 \overline D^{-*})
\]

\[
\overline a_4 = {c\over 2} (t^2_2 - p^2_2) = {{\sqrt{30}}\over 7}  
(|\overline D^-|^2 - |D^+|^2)
\]

\[
\overline a_5 = {c\over 2} (t^4_0 - p^4_0) = {6\over 7} |\overline  
D^0|^2 - {4\over 7} (|\overline D^-|^2 + |D^+|^2)
\]

\[
\overline a_6 = {c\over 2} (t^4_1 - p^4_1) = {{4\sqrt{15}}\over 7}  
{\rm Re} (\overline D^0 \overline D^{-*})
\]

\[
\overline a_7 = {c\over 2} (t^4_2 - p^4_2) = {2\over 7} \sqrt{10}  
(|\overline D^-|^2 - |D^+|^2)
\]

\noindent
The first set of 6 independent equations involves 4 moduli $|S|$,  
$|D^0|$, $|D^-|$, $|\overline D^+|$ and 3 cosines of relative phases  
$\cos(\gamma_{SD^0})$, $\cos(\gamma_{SD^-})$,  
$\cos(\gamma_{D^0D^-})$. The second set of 6 independent equations  
involves the amplitudes of opposite transversity -- 4 moduli  
$|\overline S|$, $|\overline D^0|$, $|\overline D^-|$, $|D^+|$ and 3  
cosines of their relative phases $\cos (\overline\gamma_{SD^0})$,  
$\cos(\overline\gamma_{SD^-})$ and $\cos(\overline\gamma_{D^0D^-})$.  
The two sets are entirely independent and the relative phase  
between transversity amplitudes up and down is unknown in  
measurements on transversely polarized targets.

To proceed with the analytical solution, we first find from (5.1)

\begin{equation}
|D^0|^2 = {4\over{10}} (a_1 - |S|^2) + {7\over{10}} a_5
\end{equation}

\[
|D^-|^2 = {3\over{10}} (a_1 - |S|^2) - {7\over{10}} a_5 +  
{7\over{2\sqrt{30}}} a_4
\]

\[
|\overline D^+|^2 = {3\over{10}} (a_1 - |S|^2) - {7\over{10}} a_5 -  
{7\over{2\sqrt{30}}} a_4
\]

\begin{equation}
\cos\gamma_{SD^0} = {1\over{|S| |D^0|}} (A + {1\over{2\sqrt 5}} |S|^2)
\end{equation}

\[
\cos\gamma_{SD^-} = {1\over{|S| |D^-|}} B
\]

\[
\cos\gamma_{D^0D^-} = {1\over{|D^0| |D^-|}} C
\]

\noindent
where

\begin{equation}
A = {1\over 2} \{ a_2 - {1\over{\sqrt 5}} a_1 + {1\over{2\sqrt 5}} a_5\}
\end{equation}

\[
B = {1\over 2} \{ {1\over \sqrt 2} a_3 - {1\over{2\sqrt 3}} a_6 \}
\]

\[
C = {1\over 2} \{ {7\over{4\sqrt{15}}} a_6\}
\]

\noindent
Notice that $a_7$ is not independent and does not enter in the  
above equations. Similar solutions can be derived from the second  
set (5.2) for amplitudes of opposite transversity. However we need  
one more equation in each set: one equation for $|S|^2$ in the first  
set and another one for $|\overline S|^2$ in the second set.

The additional equations are provided by the relative phases which  
are not independent:

\begin{equation}
\gamma_{SD^0} - \gamma_{SD^-} + \gamma_{D^0D^-} = (\phi_S -  
\phi_{D^0}) - (\phi_S - \phi_{D^-}) + (\phi_{D^0} - \phi_{D^-}) = 0
\end{equation}

\[
\overline\gamma_{SD^0} - \overline\gamma_{SD^-} +  
\overline\gamma_{D^0 D^-} = (\overline\phi_S - \overline\phi_{D^0})  
- (\overline\phi_S - \overline\phi_{D^-}) + (\overline\phi_{D^0} -  
\overline\phi_{D^-}) = 0
\]

\noindent
These conditions lead to nonlinear relations between the cosines:

\begin{equation}
\cos^2 (\gamma_{SD^0}) + \cos^2 (\gamma_{SD^-}) + \cos^2  
(\gamma_{D^0 D^-})
\end{equation}

\[
- 2\cos(\gamma_{SD^0}) \cos (\gamma_{SD^-}) \cos (\gamma_{D^0 D^-}) = 1
\]

\[
\cos^2 (\overline\gamma_{SD^0}) + \cos^2 (\overline\gamma_{SD^-}) +  
\cos^2 (\overline\gamma_{D^0D^-})
\]

\[
- 2\cos (\overline\gamma_{SD^0}) \cos (\overline\gamma_{SD^-}) \cos  
(\overline\gamma_{D^0 D^-}) =1
\]

\noindent
Similar relations also hold for the sines. Next we define  
combinations of observables

\begin{equation}
D = {4\over{10}} a_1 - {7\over{10}} a_5
\end{equation}

\[
E = {3\over{10}} a_1 - {7\over{20}} a_5 + {7\over{2\sqrt 3}} a_4
\]

\noindent
so that

\begin{equation}
|D^0|^2 = D - {4\over{10}} |S|^2
\end{equation}

\[
|D^-|^2 = E - {3\over{10}} |S|^2
\]

\noindent
Substituting into (5.7) first from (5.4) for the cosines and then  
from (5.9) for $|D^0|^2$ and $|D^-|^2$, we obtain a cubic equation  
for $x\equiv |S|^2$

\begin{equation}
ax^3 + bx^2 + cx + d = 0
\end{equation}

\noindent
where

\begin{equation}
a = {{27}\over{200}}
\end{equation}

\[
b = {1\over{10}} ({1\over\sqrt 5} A - 3D - {9\over 2} E)
\]

\[
c = {1\over{10}} (3A^2 + 4B^2 - 10C^2 + 2\sqrt 5 BC - 2\sqrt 5 AE + 10DE)
\]

\[
d = 2ABC - A^2 E - B^2D
\]

\noindent
Similar cubic equation can be derived for the amplitude $|\overline  
S|^2$.

Analytical expressions for the 3 roots of the cubic equation (5.10)  
are given in the Table I of Ref.~21. It is seen from the Table that  
3 real solutions exist, one of them is negative and it is rejected.  
There are in general two positive solutions for $|S|^2$ which lead  
to two solutions in the Set 1. Similarly there are two solutions in  
the Set 2 of opposite transversity. Since the two sets are  
independent there are 4 solutions for partial wave intensities

\begin{equation}
I_A (i,j) = |A(i)|^2 + |\overline A (j)|^2\ ,\ i,j=1,2
\end{equation}

The error propagation in the cubic equation and the calculation of  
errors on the moduli, cosines and partial wave intensities as well  
as the treatment of unphysical complex solutions is best handled  
using the Monte Carlo method described in detail in Ref.~24.

The determination of relative phases between natural exchange  
amplitude $D^+$ and unnatural exchange amplitudes $S$, $D^0$, $D^-$  
is described in the Appendix A.

\section{Model-independent amplitude analysis of $\pi^-  
\lowercase{p}_\uparrow \to\pi^0\pi^0 \lowercase{n}$ measured on  
polarized target with $G$-wave included.}

In the mass region where $S$-, $D$- and $G$-wave all contribute  
(expected above 1500 MeV), the measurement of $\pi^- p_\uparrow \to  
\pi^0\pi^0 n$ on polarized target will yield 13 unpolarized moments  
$t^L_M$, 13 polarized moments $p^L_M$ and 8 polarized moments  
$r^L_M$. Central to our discussion are again the moments $t^L_M$ and  
$p^L_M$ given by eqs.~(2.14) and (2.15). Using the definitions  
(2.10)--(2.13) we see again that it is useful to define two new sets  
of observables, one with the sums $t^L_M + p^L_M$ and another one  
with the differences $t^L_M - p^L_M$. With $c=\sqrt{4\pi}$ we obtain  
for the first set (sums):

\begin{equation}
a_1 = {c\over 2} (t^0_0 + p^0_0) = |S|^2 + |D^0|^2 + |D^-|^2 +  
|\overline D^+|^2 + |G^0|^2 + |G^-|^2 + |\overline G^+|^2
\end{equation}

\[
a_2 = {c\over 2} (t^2_0 + p^2_0) = \sqrt 5 \{ {2\over\sqrt 5} {\rm  
Re} (SD^{0*}) + {2\over 7} |D^0|^2 + {1\over 7} (|D^-|^2 +  
|\overline D^+|^2)
\]

\[
+ {{12}\over{7\sqrt 5}} {\rm Re} (D^0 G^{0*}) + {{2\sqrt 6}\over 7}  
[{\rm Re} (D^- G^{-*}) + {\rm Re} (\overline D^+ \overline G^{+*})]
\]

\[
+ {{20}\over{77}} |G^0|^2 + {{17}\over{77}} (|G^-|^2 + |\overline  
G^+|^2)\}
\]

\[
a_3 = {c\over 2} (t^2_1 + p^2_1) = 2\sqrt 5 \{ {2\over{\sqrt{10}}}  
{\rm Re} (SD^{-*}) + {{\sqrt 2}\over 7} {\rm Re} (D^0 D^{-*})+
\]

\[
+ {{2\sqrt 3}\over 7} {\rm Re} (D^0 G^{-*}) - {{4\sqrt  
2}\over{5\sqrt 5}} {\rm Re} (D^- G^{0*}) + {{2\sqrt{15}}\over{77}}  
{\rm Re} (G^0 G^{-*})\}
\]

\[
a_4 = {c\over 2} (t^2_2 + p^2_2) = 2\sqrt 5 \{ {1\over 7}  
\sqrt{{3\over 2}} (|D^-|^2 - |\overline D^+|^2) -
\]

\[
-{1\over 7} [{\rm Re} (D^- G^{-*}) - {\rm Re} (\overline D^+  
\overline G^{+*})] + {{5\sqrt 6}\over{77}} (|G^-|^2 - |\overline  
G^+|^2)
\]

\[
a_5 = {c\over 2} (t^4_0 + p^4_0) = \sqrt 9 \{ {2\over 7} |D^0|^2 -  
{4\over{21}} (|D^-|^2 + |\overline D^+|^2)+
\]

\[
+ {2\over 3} {\rm Re} (SG^{0*}) + {{40\sqrt 5}\over{231}} {\rm Re}  
(D^0 G^{0*}) +\]

\[
+ {{162}\over{1001}} |G^0|^2 + {{81}\over{1001}} (|G^-|^2 +  
|\overline G^+|^2) +
\]

\[
+ {{10\sqrt 2}\over{77\sqrt 3}} [{\rm Re} (D^- G^{-*}) + {\rm Re}  
(\overline D^+  \overline G^{+*})]\}
\]

\[
a_6 = {c\over 2} (t^4_1 + p^4_1) = 2\sqrt 9 \{ {2\over 7}  
\sqrt{{5\over 3}} {\rm Re} (D^0 D^{-*}) + {{\sqrt 2}\over 3} {\rm  
Re} (SG^{-*})
\]

\[
+ {{17\sqrt{10}}\over{231}} {\rm Re} (D^0 G^{-*}) +  
{{10}\over{77\sqrt 3}} {\rm Re} (D^- G^{0*}) + {{81\sqrt  
2}\over{1001}} {\rm Re} (G^0 G^{-*})\}
\]

\[
a_7 = {c\over 2} (t^4_2 + p^4_2) = 2\sqrt 9 \{  
{{\sqrt{10}}\over{21}} (|D^-|^2 - |\overline D^+|^2 ) +
\]

\[
+ {{6\sqrt{15}}\over{154}} [{\rm Re} (D^- G^{-*}) - {\rm Re}  
(\overline D^+ \overline G^{+*}) + {{27\sqrt{10}}\over{1001}}  
(|G^-|^2 - |\overline G^+|^2)\}
\]

\[
a_8 = {c\over 2} (t^6_0 + p^6_0) = {{\sqrt{13}}\over{143}} \{  
30\sqrt 5 {\rm Re} (D^0 G^{0*}) - 20\sqrt 6 [{\rm Re} (D^- G^{-*})+
\]

\[
+ {\rm Re} (\overline D^+ \overline G^{+*})] + 20 |G^0|^2 -  
(|G^-|^2 + |\overline G^+|^2)\}
\]

\[
a_9 = {c\over 2} (t^6_1 + p^6_1) = {{2\sqrt{13}}\over{143}} \{ 10  
\sqrt{21} {\rm Re} (D^0 G^{-*}) + {{10\sqrt{35}}\over{\sqrt 2}} {\rm  
Re} (D^- G^{0*})
\]

\[
+ 2\sqrt{105} {\rm Re} (G^0 G^{-*})\}
\]

\[
a_{10} = {c\over 2} (t^6_2 + p^6_2) = {{2\sqrt{13}}\over{143}} \{  
4\sqrt{70} [{\rm Re} (D^- G^{-*}) - {\rm Re} (\overline D^+  
\overline G^{+*})]
\]

\[
+ \sqrt{105} (|G^-|^2 - |\overline G^+|^2)\}
\]

\[
a_{11} = {c\over 2} (t^8_0 + p^8_0) = {{\sqrt{17}}\over{2431}} \{  
490 |G^0|^2 - 392 (|G^-|^2 + |\overline G^+|^2)\}
\]

\[
a_{12} = {c\over 2} (t^8_1 + p^8_1) = {{2\sqrt{17}}\over{2431}} \{  
294 \sqrt 5 {\rm Re} (G^0 G^{-*})\}
\]

\[
a_{13} = {c\over 2} (t^8_2 + p^8_2) = {{2\sqrt{17}}\over{2431}} \{  
42\sqrt{35} (|G^-|^2 - |\overline G^+|^2)\}
\]

\noindent
The second set of observables $\overline a_i, i=1,2,\ldots,13$ is  
formed similarly by the differences $t^L_M - p^L_M$. It has the same  
form as set 1 but involves the amplitudes of opposite transversity.

The first set $a_i, i=1,\ldots,13$ involves 7 moduli

\begin{equation}
|S|, |D^0|, |D^-|, |\overline D^+|, |G^0|, |G^-|, |\overline G^+|,
\end{equation}

\noindent
10 cosines of relative phases between unnatural amplitudes

\begin{equation}
\cos (\gamma_{SD^0}), \cos(\gamma_{SD^-}), \cos (\gamma_{SG^0}),  
\cos(\gamma_{SG^-})
\end{equation}

\begin{equation}
\cos (\gamma_{D^0D^-}), \cos(\gamma_{D^0G^0}), \cos(\gamma_{D^0G^-})
\end{equation}

\begin{equation}
\cos(\gamma_{D^-G^0}), \cos(\gamma_{D^- G^-}), \cos(\gamma_{G^0G^-})
\end{equation}

\noindent
and one cosine of relative phase between the two natural amplitudes

\begin{equation}
\cos(\overline\gamma_{D^+G^+})
\end{equation}

\noindent
The second set $\overline a_i, i=1,\ldots ,13$ involves the same  
amplitudes but of opposite transversity. We will now show that the  
cosines (6.4) and (6.5) can be expressed in terms of cosines (6.3).  
For instance, we can write

\begin{equation}
\gamma_{D^0 D^-} = \phi_{D^0} - \phi_{D^-} = (\phi_S - \phi_{D^-})  
- (\phi_S - \phi_{D^0}) = \gamma_{SD^-} - \gamma_{SD^0}
\end{equation}

\noindent
Hence

\[
\cos\gamma_{D^0D^-} = \cos\gamma_{SD^0} \cos\gamma_{SD^-} +  
\sin\gamma_{SD^0} \sin\gamma_{SD^-}
\]

\noindent
Since the signs of the sines $\sin\gamma_{SD^0}$ and  
$\sin\gamma_{SD^-}$ are not known, we write

\begin{equation}
\sin\gamma_{SD^0} = \epsilon_{SD^0} |\sin\gamma_{SD^0}|\ ,\  
\sin\gamma_{SD^-} = \epsilon_{SD^-} |\sin\gamma_{SD^-}|
\end{equation}

\noindent
Then

\begin{equation}
\cos\gamma_{D^0D^-} = \cos\gamma_{SD^0} \cos\gamma_{SD^-} +  
\epsilon_{D^0D^-} \sqrt{(1-\cos^2\gamma_{SD^0})  
(1-\cos^2\gamma_{SD^-})}
\end{equation}

\noindent
where $\epsilon_{D^0D^-} = \pm 1$ is the sign ambiguity. The  
remaining cosines in (6.4) and (6.5) can be written in the form  
similar to (6.9) with their own sign ambiguities. The sign  
ambiguities of all cosines (6.4) and (6.5) can be written in terms  
of sign ambiguities $\epsilon_{SD^0}, \epsilon_{SD^-},  
\epsilon_{SG^0}, \epsilon_{SG^-}$ corresponding to the sines  
$\sin\gamma_{SD^0}, \sin\gamma_{SD^-}, \sin\gamma_{SG^0},  
\sin\gamma_{SG^-}$. We can write

\begin{equation}
\epsilon_{D^0D^-} = \epsilon_{SD^0} \epsilon_{SD^-}
\end{equation}

\[
\epsilon_{D^0G^0} = \epsilon_{SD^0} \epsilon_{SG^0}
\]

\[
\epsilon_{D^0G^-} = \epsilon_{SD^0} \epsilon_{SG^-}
\]

\begin{equation}
\epsilon_{D^-G^0} = \epsilon_{SD^-} \epsilon_{SG^0}
\end{equation}

\[
\epsilon_{D^-G^-} = \epsilon_{SD^-} \epsilon_{SG^-}
\]

\[
\epsilon_{G^0G^-} = \epsilon_{SG^0} \epsilon_{SG^-}
\]

\noindent
First we notice that reversal of all signs $\epsilon_{SD^0},  
\epsilon_{SD^-}, \epsilon_{SG^0}$, and $\epsilon_{SG^-}$ yields the  
same sign ambiguities (6.10) and (6.11). Next we notice that the  
sign ambiguities (6.11) of cosines (6.5) are uniquely determined by  
the sign ambiguities (6.10) for cosines (6.4). Only sign ambiguities  
(6.10) are independent and there is 8 sign combinations (6.10). The  
following table lists all 8 allowed sets of sign ambiguities of  
cosines (6.4) and (6.4).

\begin{quasitable}
\begin{tabular}{lcccccccc}
&1 &2 &3 &4 &5 &5 &7 &8 \\
\tableline
$\epsilon_{D^0D^-}$ &$+$ &$-$ &$+$ &$+$ &$-$ &$-$ &$+$ &$-$ \\
$\epsilon_{D^0G^0}$ &$+$ &$+$ &$-$ &$+$ &$-$ &$+$ &$-$ &$-$ \\
$\epsilon_{D^0G^-}$ &$+$ &$+$ &$+$ &$-$ &$+$ &$-$ &$-$ &$-$ \\
$\epsilon_{D^-G^0}$ &$+$ &$-$ &$-$ &$+$ &$+$ &$-$ &$-$ &$+$ \\
$\epsilon_{D^-G^-}$ &$+$ &$-$ &$+$ &$-$ &$-$ &$+$ &$-$ &$+$ \\
$\epsilon_{G^0G^-}$ &$+$ &$+$ &$-$ &$-$ &$-$ &$-$ &$+$ &$+$
\end{tabular}
\end{quasitable}

Using expressions like (6.9) for cosines (6.4) and (6.5), we have  
12 unknowns in each nonlinear set of 13 equations $a_i,  
i=1,2,\ldots,13$ with one choice of sign ambiguities for cosines  
(6.4) and (6.5) from the above Table. The nonlinear set can be  
solved numerically or by $\chi^2$ method. In each $(m,t)$ bin we  
thus have 8 solutions for moduli (6.2) and cosines (6.3)--(6.5), and  
8 solutions for amplitudes of opposite transversity from the set  
$\overline a_i, i=1,2,\ldots,13$. Since each solution is uniquely  
labeled by the choice of sign ambiguities, there is no problem  
linking solutions in neighbouring $(m,t)$ bins.

Since the 8 solutions from the first set $a_i, i=1,2,\ldots,13$ are  
independent from the 8 solutions from the second set $\overline  
a_i, i=1,2,\ldots,13$, there will be 64-fold ambiguity in the  
partial wave intensities. For $A=S,D^0,D^-, D^+, G^0, G^-, G^+$ we  
can write

\begin{equation}
I_A (i,j) = |A(i)|^2 + |\overline A (j)|^2\ ,\ i,j=1,2,\ldots,8
\end{equation}

We now will discuss constraints on the moments that should be taken  
into account at the time of fitting maximum likelihood function  
${\cal L}$ to the observed angular distribution in the process of  
constrained optimization.

The observables $a_i$ and $\overline a_i, i=1,2,\ldots,13$ are not  
all linearly independent. In fact one finds two  
relations\cite{chung96b}

\begin{equation}
8\sqrt{14} a_4 - 4\sqrt{42} a_7 + {{91}\over{\sqrt{13}}} a_{10} -  
{{119}\over{2}} \sqrt{{3\over{17}}} a_{13} = 0
\end{equation}

\[
8\sqrt{14} \overline a_4 - 4\sqrt{42} \overline a_7 +  
{{91}\over{\sqrt{13}}} \overline a_{10} - {{119}\over{2}}  
\sqrt{{3\over{17}}} \overline a_{13} = 0
\]

\noindent
By adding and subtracting the last two equations we get the same  
relationship for corresponding moments $t^L_M$ and $p^L_M$:

\begin{equation}
8\sqrt{14} t^2_2 - 4\sqrt{42} t^4_2 + {{91}\over{\sqrt{13}}} t^6_2  
- {{119}\over 2} \sqrt{{3\over{17}}} t^8_2 = 0
\end{equation}

 \[
8\sqrt{14} p^2_2 - 4\sqrt{42} p^4_2 + {{91}\over{\sqrt{13}}} p^6_2  
- {{119}\over 2} \sqrt{{3\over{17}}} p^8_2 = 0
\]

\noindent
Additional constraints can be obtained by solving for $|G^-|^2 +  
|\overline G^+|^2$ from $a_{11}$ and substituting into $a_1$.  
Proceeding in the same way also for $|\overline G^-|^2 + |G^+|^2$  
from $\overline a_{11}$ and substituting into $\overline a_1$, we  
get

\begin{equation}
a_1 + {{2431}\over{392\sqrt{17}}} a_{11} > 0\ ,\ \overline a_1 +  
{{2431}\over{392\sqrt{17}}} \overline a_{11} > 0
\end{equation}

\noindent
By adding the two inequalities we get

\begin{equation}
t^0_0 + {{2431}\over{392\sqrt{17}}} t^8_0 > 0
\end{equation}

\noindent
The constraints (6.13) and (6.14), or (6.15) and (6.16), are  
self-consistency constraints which follow from the assumption that  
only $S$-, $D$- and $G$-waves contribute. These constraints should  
be imposed on the maximum likelihood function during the fit to the  
observed angular distribution. We then deal with constrained  
optimization\cite{ref39,gill81,rao84}. A program MINOS 5.0 has been  
developed at Stanford University for constrained optimization with  
equalities and inequalities constraints\cite{murtagh83}.

\section{Summary.}

The dependence of hadronic reactions on nucleon spin is now a  
well-established experimental fact. The measurements of reactions  
$\pi^- p \to \pi^-\pi^+ n$ and $\pi^+ n \to \pi^+ \pi^- p$ on  
polarized targets at CERN found a strong dependence of pion  
production amplitudes on nucleon spin. The assumption thtat pion  
production amplitudes are independent of nucleon spin is in direct  
conflict with these experimental findings. The analyses of $\pi^- p  
\to \pi^0\pi^0 n$ data based on this assumption thus are not  
sufficient and may not be fully reliable.

We have shown in Section III that unpolarized data provide model  
independent information only on the spin averaged partial wave  
intensities and cosines of three interference phases. To obtain  
information about the production amplitudes, measurements of $\pi^-  
p \to \pi^0 \pi^0 n$ on polarized target are necessary. We have  
shown in Sections V and VI how to perform model independent  
amplitude analysis of $\pi^- p \to \pi^0\pi^0 n$ measured on  
polarized targets. Model independent analysis is possible in the  
mass region where only $S$- and $D$-wave amplitudes contribute, as  
well as in the mass region where also $G$-wave amplitudes  
contribute. Our only assumption was that amplitudes with dimeson  
helicity $\lambda \ge 2$ do not significantly contribute to the  
$\pi^0\pi^0$ production. This assumption is supported by the  
available data.

On this basis we propose that high statistics measurements of  
$\pi^- p \to \pi^0\pi^0 n$ and $\pi^- p \to \eta\eta n$ be made at  
BNL Multiparticle Spectrometer and at IHEP in Protvino and that  
model independent amplitude analysis of these reactions be  
performed. We note that this amplitude analysis will require the  
unpolarized moments $t^L_M$ which should be determined from the data  
on unpolarized targets in the same $t$-bins.

We suggest that the extensions of GAMS and BNL E852 program to  
measurements on polarized targets will significantly contribute to  
new developments of hadron spectroscopy on the level of spin  
dependent production amplitudes and to our understanding of hadron  
dynamics.

\acknowledgements
I wish to thank B.B.~Brabson, J.~Gunter, A.A.~Kondashov,  
Yu.~D.~Prokoshkin and S.A.~Sadovsky for stimulating e-mail  
correspondence concerning the E852 and GAMS experiments on $\pi^- p  
\to \pi^0\pi^0 n$. This work was supported by Fonds pour la  
Formation de Chercheurs et l'Aide \`a la Recherche (FCAR),  
Minist\`ere de l'Education du Qu\'ebec, Canada.

\appendix
\section*{Calculation of phases $\gamma_{D^+S}$ and  
$\overline\gamma_{D^+S}$}

In this appendix we solve Eqs.~(2.16) for the helicity frame  
invariant phases $\gamma_{D^+S} = \phi_{D^+} - \phi_S$ and  
$\overline\gamma_{D^+S} = \overline\phi_{D^+} - \overline\phi_S$.  
Other phases in (2.16) are then expressed in terms of these phases  
and the phases (5.4):

\begin{equation}
\gamma_{D^+D^0} = \phi_{D^+} - \phi_{D^0} = (\phi_{D^+} - \phi_S) +  
(\phi_{S} - \phi_{D^0}) = \gamma_{D^+ S} - \gamma_{D^0S}
\end{equation}

\[
\gamma_{D^+D^-} = \phi_{D^+} - \phi_{D^-} = (\phi_{D^+} - \phi_S) +  
(\phi_{S} - \phi_{D^-}) = \gamma_{D^+ S} - \gamma_{D^-S}
\]

\noindent
with similar relations for $\overline\gamma_{D^+ D^0}$ and  
$\overline\gamma_{D^+ D^-}$. The system of equations (2.16) can then  
be written as

\begin{equation}
b_1 = {{7\sqrt{4\pi}}\over{2\sqrt{30}}} r^2_2 = |D^+| |D^-|  
\cos\gamma_{D^+ D^-} - |\overline D^+| |\overline D^-|  
\cos\overline\gamma_{D^+ D^-}
\end{equation}

\[
b_2 = {{7\sqrt{4\pi}}\over{4\sqrt{15}}} r^4_1 = |D^+| |D^0|  
\cos\gamma_{D^+ D^0} - |\overline D^+| |\overline D^0|  
\cos\overline\gamma_{D^+ D^0}
\]

\[
b_3 = {{\sqrt{4\pi}}\over{2\sqrt 2}} r^2_1 - \sqrt{{5\over 7}} b_2  
= |D^+| |S| \cos\gamma_{D^+S} - |\overline D^+| |\overline S| \cos  
\overline\gamma_{D^+S}
\]

\noindent
>From $b_3$ we obtain

\begin{equation}
\cos\overline\gamma_{D^+S} = {{|D^+| |S| \cos\gamma_{D^+S} -  
b_3}\over{|\overline D^+| |\overline S|}}
\end{equation}

\noindent
Using (A1) we obtain from $b_2$

\begin{equation}
\sin\overline\gamma_{D^+S} = - \cos\overline\gamma_{D^+S}  
(\cos\overline\gamma_{D^0S}/\sin\overline\gamma_{D^0S}) +
\end{equation}

\[
+ {{b_2 - |D^+| |D^0| (\cos\gamma_{D^+S} \cos\gamma_{D^0S} +  
\sin\gamma_{D^+S} \sin\gamma_{D^0S})}\over{|\overline D^+|  
|\overline D^0| \sin\overline\gamma_{D^0S}}}
\]

\noindent
We now define

\begin{equation}
c_1\equiv |D^0| |S| \sin\gamma_{D^0S} = \epsilon_1 \sqrt{|D^0|^2  
|S|^2 - (A + {1\over{2\sqrt 5}} |S|^2)^2}
\end{equation}

\[
c_2\equiv |D^-| |S| \sin\gamma_{D^-S} = \epsilon_2 \sqrt{|D^-|^2  
|S|^2 - B^2}
\]

\[
c_3 = |D^-| |D^0| \sin\gamma_{D^-D^0} = \epsilon_3 \sqrt{|D^-|^2  
|D^0|^2 - C^2}
\]

\noindent
where $\epsilon_k = \pm 1$,$ k=1,2,3$ is the ambiguity sign of the  
sines. The $c_3$ and the sign $\epsilon_3$ are not independent of  
$c_1$ and $c_2$:

\begin{equation}
|S|^2 c_3 = (A + {1\over{2\sqrt 5}} |S|^2) c_2 - Bc_1
\end{equation}

\noindent
Similarly we define $\overline c_1$, $\overline c_2$ and $\overline  
c_3$ for amplitudes of opposite transversity. Substituting for  
$\cos\overline\gamma_{D^+S}$ and $\sin\overline\gamma_{D^+S}$ from  
(A3) and (A4) in the equation for $b_1$ and using the above  
definitions for $c_k$, $\overline c_k$, $k=1,2,3$, we obtain

\begin{equation}
(b_1 \overline c_1 + b_2 \overline c_2 + b_3 \overline c_3)  
|\overline S|^2 |S| =
\end{equation}

\[
\sin\gamma_{D^+ S} |D^+| |\overline S|^2 (c_1 \overline c_2 +  
\overline c_1 c_2) +
\]

\[
+ \cos\gamma_{D^+S} |D^+| \{ \overline c_1 (B |\overline S|^2 -  
\overline B |S|^2) +
\]

\[
+ \overline c_2 [(A + {1\over{2\sqrt 5}} |S|^2) |\overline S|^2 +  
(\overline A + {1\over{2\sqrt 5}} |\overline S|^2) |S|^2]\}
\]

\noindent
Define

\begin{equation}
d = {{b_1 \overline c_1 + b_2 \overline c_2 + b_3 \overline  
c_3}\over{c_1 \overline c_2 + \overline c_1 c_2}}  
({{|S|}\over{|D^+|}})
\end{equation}

\[
\tan\alpha = \{ \overline c_1 (B |\overline S|^2 - \overline B  
|S|^2) + \overline c_2 [(A + {1\over{2\sqrt 5}} |S|^2) |\overline  
S|^2 +
\]

\[
+ (\overline A + {1\over{2\sqrt 5}} |\overline S|^2) |S|^2 ]\} /  
(c_1 \overline c_2 + \overline c_1 c_2) |\overline S|^2
\]

\noindent
With this notation (A6) takes the form

\begin{equation}
\sin\gamma_{D^+S} + \cos\gamma_{D^+S} \tan\alpha = d
\end{equation}

\noindent
Its solution is

\begin{equation}
\cos\gamma_{D^+S} = {1\over{1 + \tan^2\alpha}} \{ d\tan\alpha \pm  
\sqrt{1 + \tan^2\alpha - d^2}\}
\end{equation}

\[
\sin\gamma_{D^+S} = {1\over{1+\tan^2\alpha}} \{ d \mp \tan\alpha  
\sqrt{1 + \tan^2\alpha - d^2} \}
\]

\noindent
Using (A10) we obtain $\cos\overline\gamma_{D^+S}$ and  
$\sin\overline\gamma_{D^+S}$ from (A3) and (A4).

There are four combinations of solutions for moduli $|A|^2$,  
$|\overline A|^2$, $A=S,D^0, D^-, D^+$ entering the calculation of  
$d$ and $\tan\alpha$. In addition each such combination is  
accompanied by the fourfold sign ambiguity from the undetermined  
signs $\epsilon_k$ and $\overline\epsilon_k$, $k=1,2$. This 16-fold  
ambiguity increases to 32-fold ambiguity due to sign ambiguity in  
(A10).

The solvability of (A9) imposes a nonlinear constraint on data and  
on the solution for moduli squared

\begin{equation}
d^2 - 1 \le \tan^2\alpha
\end{equation}

\noindent
Additional constraints follow from the requirement that cosines and  
sines of $\gamma_{D^+S}$ and $\overline\gamma_{D^+S}$ have physical  
values. In principle, these constraints could reduce the overall  
ambiguity of solution (A10).

\begin{figure}
\caption{Polarized target asymmetry in reactions $\pi^- p \to \pi^-  
\pi^+ n$ and $\pi^+ n \to \pi^+ \pi^- p$. The assumption that the  
pion production amplitudes do not depend on nucleon spin predicts  
that polarized target asymmetry be zero.}\label{fig1}
\end{figure}

\begin{figure}
\caption{The ratio of amplitudes with recoil nucleon transversity  
``down'' and ``up'' with dimeson helicity $\lambda=0$ in $\pi^- p  
\to \pi^-\pi^+ n$ at 17.2 GeV/c and $-t = 0.005 - 0.2$ (GeV/c)$^2$.  
The assumption that the pion production amplitudes do not depend on  
nucleon spin predicts that all ratios be equal to 1. The deviation  
from unity shows the strength of dependence of production amplitudes  
on nucleon spin. Based on Fig.~6 of Ref.~14.}\label{fig2}
\end{figure}

\begin{figure}
\caption{Mass dependence of unnormalized amplitudes $|\overline  
S|^2\Sigma$ and $|S|^2\Sigma$ measured in $\pi^- p_\uparrow \to  
\pi^- \pi^+ n$ at 17.2 GeV/c at $-t = 0.005 - 0.20$ (GeV/c)$^2$  
using the Monte Carlo method for amplitude analysis (Ref.~24). Both  
solutions for the transversity ``up'' amplitude $|\overline  
S|^2\Sigma$ resonate while the transversity ``down'' amplitude  
$|S|^2\Sigma$ is nonresonating in both solutions.}\label{fig3}
\end{figure}

\begin{figure}
\caption{Four solutions for the $S$-wave intensity $I_S$ measured  
in the reaction $\pi^- p_\uparrow \to \pi^-\pi^+ n$ at 17.2 GeV/c  
and $-t = 0.005 - 0.20$ GeV/c using Monte Carlo method for amplitude  
analysis (Ref.~24). Although both solutions for amplitude  
$|\overline S|^2\Sigma$ resonate, the intensity $I_S(2,2)$ appears  
nonresonating.}\label{fig4}
\end{figure}

\begin{figure}
\caption{Definition of the coordinate systems used to describe the  
target polarization ${\vec P}$ and the decay of the dimeson  
$\pi^0\pi^0$ system.}\label{fig5}
\end{figure}

\end{document}